**APL Special Topic Guest Editorial**

**Mesoscopic magnetic systems: from fundamental properties to devices**

Laura J. Heyderman[1,2], Julie Grollier[3], Christopher H. Marrows[4], Paolo Vavassori[5], Dirk Grundler[6], Denys Makarov[7], Salvador Pané[8]

[1]Laboratory for Mesoscopic Systems, Department of Materials, ETH Zurich, 8093 Zurich, Switzerland

[2]Laboratory for Multiscale Materials Experiments, Paul Scherrer Institute, 5232 Villigen PSI, Switzerland

[3]Unité Mixte de Physique CNRS, Thales, Université Paris-Saclay, 91767 Palaiseau, France

[4]School of Physics & Astronomy, University of Leeds, Leeds LS2 9JT, United Kingdom

[5]CIC nanoGUNE BRTA, 20018 San Sebastian and IKERBASQUE, Basque Foundation for Science, 48009 Bilbao, Spain

[6]Laboratory of Nanoscale Magnetic Materials and Magnonics, Institute of Materials and Institute of Electrical and Micro Engineering, Ecole Polytechnique Fédérale de Lausanne (EPFL), 1015 Lausanne, Switzerland

[7]Helmholtz-Zentrum Dresden-Rossendorf e.V., Institute of Ion Beam Physics and Materials Research, 01328 Dresden, Germany

[8]Multi-Scale Robotics Lab, Institute of Robotics and Intelligent Systems, ETH Zürich, Tannenstrasse 3, CH 8092, Zürich, Switzerland

**INTRODUCTION**

Research into mesoscopic magnetic systems, which incorporate magnetic elements with dimensions ranging from a few nm up to a few 10s of $\mu$m, has been spurred on by the developments in their fabrication, characterisation, and control. Electron beam and optical lithography are key methods for their fabrication because they can be used to create tailor-made planar arrangements of elements or more complex layered structures. These can be further transformed into high quality three-dimensional mesoscale architectures using strain engineering based origami approaches. Further methods include fabrication of magnetic nano- and micro-objects on curved templates prepared using ion beam erosion techniques or via self-assembly of non-magnetic particles, which does not require expensive equipment. In addition, direct write methods such as focused electron or ion beam induced deposition, along with two photon laser lithography, provide a means to create geometrically curved and three dimensional structures. The characterisation of the magnetic states can be performed with electrical measurements or using various magnetic microscopy, tomography, and scattering methods. These characterisation methods can then be used to observe the behaviour of complex magnetic textures including magnetic domain walls, vortices, Bloch points, and skyrmions. The control of the magnetic states, which is important for applications, can be achieved with a variety of external stimuli including magnetic or electric fields, spin currents, strain, photons and heat. This results in magnetization dynamics that can happen at timescales ranging from minutes down to less than a picosecond. Mesoscopic magnetic systems and their magnetic configurations are therefore not only of fundamental interest, but have the potential to be implemented in a wide range of device applications including

ferro-, ferri- and antiferromagnetic spintronics, computation, magnonics, mechanically flexible and printable human-machine interfaces, electromobility, and medical applications including small-scale machinery.

This Special Topic on mesoscopic magnetic systems includes, but is not restricted to, articles in the areas of current interest outlined below, which include experiments, theory and simulations in this fascinating field. We hope to have captured the variety of exciting areas in this Special Topic and that this will provide inspiration for the future, either in terms of fundamental science and discovery, or for more applied work to create novel devices.

**Artificial Spin Ice**

Since the original work on artificial spin ice by Peter Schiffer and his group [1], there has been a tremendous increase in interest in the field of artificial spin ice [2]. This began with the aim to mimic the behaviour of spins in the rare earth titanate pyrochlores with the magnetic configurations in arrays of dipolar coupled nanomagnets arranged on the square and kagome lattices. A broad variety of fundamental phenomena have now been investigated including frustration, emergent magnetic monopoles and phase transitions and, in this Special Topic, Peter Schiffer and Cristiano Nisoli provide an overview of the field with current areas of interest and perspectives towards the future [3]. In further articles, it is shown how the detailed arrangement of the nanomagnets play an important role in defining the collective properties [4], both the static and dynamic, including magnetisation reversal and spin wave modes [5,6]. In addition, the importance of the applied magnetic field orientation and magnitude is highlighted [7] and, in an artificial spin ice made up of a network of nanowires, it is shown how the dynamic response is particularly sensitive to the vertex configurations and positions of domain walls [8]. Cheenikundil and Hertel [9] show with remarkable micromagnetic simulations what happens when such a wire network is transformed into a three-dimensional buckyball where the magnetic states are very sensitive to the direction of applied magnetic field. Another fertile area for future research is hybrid systems, combining artificial spin ice with another class of materials such as superconductors [10]. Finally, Rougemaille and Canals highlight the importance of the magnetic structure factor in interpreting data [11] and Slöetjes et al. indicate the shortcomings of considering the nanomagnets as a macrospin and that, for many applications, the internal magnetic structure in the nanomagnets that make up the artificial spin ices should be considered [12].

**Computation and Spintronics with Nanoscale Magnets**

Computation requires the transformation, movement and storage of information and spintronics is promising in this area because it brings reliable and low energy methods to perform these three operations though the multifunctionality of nanoscale magnets. The spin-torque random access memory, now available in the production lines of major foundries, offers massive amounts of fast, low energy, non-volatile memory. In addition, non-linear magnetization dynamics through the effect of magnetic fields, spin currents, voltages or optical pulses in confined magnetic dots transforms information by switching, tilting, deforming or exciting the magnetization locally. Finally, magnons and solitons such as magnetic domain walls or skyrmions, which can be controllably displaced in magnetic films, constitute moving bits of information. Spintronics is therefore promising for a wide range of computing applications. Magnetization switching naturally supports Boolean logic operations, and Walker et al. [13] and Baumgaertl et al. [14] unravel key operations for skyrmion-based and magnon-based logic, respectively. In addition, Yang et al. introduce ferromagnetic nanodots in the barrier of superconducting tunnel junctions to understand the manipulation of magnetic information by the Kondo effect [15].

Magnetization dynamics and soliton propagation are also enablers of neuromorphic computing, as they provide a means to implement complex synapses and neurons at the nanoscale. In this Special Topic, Welbourne et al. show that voltage-controlled superparamagnetic ensembles are low-energy platforms for hardware-based reservoir computing [16]. Liu et al. propose to use domain walls in a notched track to mimic artificial synapses with high-quality weight updates [17]. Temple et al. build a memristor synapse harnessing the metamagnetic phase transition in the B2-ordered alloy FeRh [18]. Williame et al. [19] and Shen et al. [20] show through simulations that magnetic domain walls emulate Mackey-Glass and Duffing oscillators respectively, both useful for emulating the rhythmic features and synchronization properties of neurons. Zeng et al. simulate a neuromorphic network of six oscillators coupled through a chiral coupling [21]. Jenkins et al. demonstrate experimentally the analog nature of locked spin-torque nano-oscillators, promising for the design of oscillator-based neural networks [22]. Finally, Finocchio et al. underline the interest of spintronic diodes for emulating both synapses and neurons, as well as integrating sensing functions in a computing network [23]. A future challenge for all computing applications is the need for engineered barrier materials and magnetic tunnel junctions with a high Tunnel Magneto-Resistance and a small resistance-area product as considered in Ref. [24].

**Mesoscopic Topological Features: Skyrmions, Bloch Points and Vortices**

Topology is one of the dominant themes of contemporary condensed matter physics, with the topology of the physical, electronic, or magnetic structure capable of affecting the functional properties of a material in profound ways.

Here we focus on the magnetic structure and how the topology of the magnetic material itself plays a role in many of the topics covered by this Special Topic. For instance, a cylindrical wire has a different topology to a tube of the same diameter, giving rise to different domain wall types [25]. These domain walls are themselves topological spin textures and can contain topological objects like Bloch points [25,26]. The effects of curvature on magnetism are reviewed in more detail below. A key feature of artificial spin ices is that network topologies can be designed [2,3], with the buckyball reported by Cheenikundil and Hertel being an exciting three-dimensional example [9]. The field of artificial spin ice is reviewed above in more detail.

A very active topic of study in mesoscopic magnetism is that of magnetic skyrmions that are topologically non-trivial spin textures with particle-like properties, of which vortices are precursors. In this Special Topic, Zeng et al. describe using an interfacial Dzyaloshinkii-Moriya interaction (DMI) to control the chirality of an imprinted vortex with respect to a fixed polarity defined by an imprinting out-of-plane layer, and how these vortices synchronise when oscillating [21]. As with more and more mesoscopic magnetic systems these have applications in neuromorphic computing (discussed above). Complex three-dimensional spin-textures involving both vortices and antivortices were observed in thick Permalloy nanoelements by Han et al. [27], with a topology dependent stability.

The first step in studying skyrmions is to create one, which is a process known as nucleation. In this Special Topic there are descriptions of methods for skyrmions creation using focussing of spin waves [28], self-nucleation in multilayer dots [29], and all-optical nucleation (and annihilation) of skyrmions [30]. As a next step, one can probe their dynamics. A methodology is reported by van Elst et al. [31] on how to accurately quantify higher order terms in the spin-orbit torque (SOT) effect that can be especially useful for modelling both non-uniform magnetic textures such as skyrmions and their current-induced magnetization switching. Element-specific x-ray microscopy is used by Finizio et al. [32] to visualize the SOT-induced magnetization canting. This technique will allow for the experimental quantification of SOT in multilayers such as the FeTb/Pt/FeTb trilayer system presented by Dong et al. [33] that gives rise to a three-dimensional spintronic device hosting exponentially increased magnetization states. Returning again to magnetism in curved geometries, Carvalho-Santos et al. describe skyrmion propagation along curved racetracks, and show that a curved region provides a barrier to skyrmion motion [34]. Skyrmion dynamics in synthetic antiferromagnets was simulated by Qiu et al., who found a breathing mode in response to an oscillating magnetic anisotropy [35]. This kind of change in anisotropy can be driven by an electric field, and Walker et al. describe this as a means to clock a skyrmion-based logic architecture [13]. Last of all, the community is beginning to look beyond skyrmions to other kinds of topological spin textures: Zhang et al. report their simulations of a spin texture in an in-plane magnetised film for which they coin the name 'frustrated bimeronium' [36].

**Magnetoplasmonics/Magnetophotonics**

Plasmonics, namely the study of collective electromagnetic excitations of nanostructured materials and of their formidable ability to couple free-space electromagnetic radiation as well as to strongly localize and enhance it at the nanoscale, has become a burgeoning research field with applications spanning energy harvesting, telecommunications, and sensing. To reach new functionalities, the combination of conventional plasmonic materials, typically noble metals, with other materials properties has become increasingly appealing. The combination with magnetic, both ferromagnetic and ferrimagnetic, materials that display non-reciprocal magneto-optical (MO) properties is widely explored [37]. Such systems exhibit simultaneously magnetic and plasmonic properties that has led to the concept of magnetoplasmonics. They display enhanced MO effects that are exploited in metasurfaces and metacrystals aimed at the size reduction of key photonic devices based on non-reciprocal propagation of light. Active plasmonics, namely the active control of surface plasmon resonance, is also a growing and challenging subfield of plasmonics. Again, the use of magnetic materials is envisaged as a means to achieve the active control of the plasmonic response via an applied magnetic field. In this Special Topic, Gaspar Armelles and Alfonso Cebollada show how to achieve the magnetic modulation of plasmon resonances in the mid-infrared to the far-infrared spectral ranges using a FeNi metamaterial [38]. Likewise, but this time without using metallic materials, Kil-Song Song et al. explore theoretically how optical transparency of planar semiconductor interfaces can be switched on and off by a magnetic field [39]. Magnetoplasmonics also provides a way to create chiroptical metasurfaces where the chiral light transmission is controlled by the applied magnetic field [40] as shown in this Special Topic by Gaia Petrucci et al. [41].

**Magnetization Dynamics and Spin Waves in Nanomagnets, Magnonic Crystals and Circuits**

Magnonics has generated steadily increasing interest during the last decade [42]. It aims at the functionalization of spin waves (magnons) for applications in information technology, computation, and on-chip signal processing at microwave frequencies from GHz to THz. This might be achieved for instance by means of radiofrequency filters based on mesoscopic spin-wave cavities [43] or circulators enabled by chiral spin-wave routing across two stacked ferromagnetic layers without interlayer exchange coupling [44]. Magnetization dynamics in synthetic antiferromagnets, i.e., acoustic and optical magnons in layers with interlayer-exchange coupling, are considered theoretically by Dai et al. [45] and Jeffrey et al. [46]. Aiming at magnonic platforms for devices exploiting hybridized quantum excitations, they predict strong magnon-magnon coupling in tilted fields and a suppressed exceptional point, respectively. Exploring propagating spin waves in one-dimensional magnonic crystals, Grachev et al. demonstrate a strain-controlled band structure which enables reprogrammable frequency-selective multiplexers [47], while Baumgaertl et al. in Ref. [14] demonstrate a magnon phase inverter exploiting an engineered bound state of an individual nanomagnet. In Ref. [48], advanced x-ray imaging allowed Groß et al. to unravel an unexpected band structure transformation in a two-dimensional magnonic crystal. Individual straight and coupled curved magnon conduits based on nanostructured CoFeB and yttrium iron garnet (YIG) are explored theoretically and experimentally in Refs. [49-51]. Based on low-damping YIG, researchers have discovered shape-induced magnon non-reciprocity [50] and realize the quantum-classical analogy of a dark state in nanomagnonics [51]. Exchange dominated magnons in nanostructured magnets provide a particular advantage in that they exhibit isotropic properties and enable magnon transport in complex three-dimensional (3D) device architectures. Non-planar devices leading towards innovative 3D magnonics are realized by the ferromagnetic nanovolcanos of Ref. [52], nanotubes with non-reciprocal magnon transport of Ref. [53] and the surface-corrugated ferromagnets of Ref. [54] which allow for fast magnons in multi-directional circuits at zero magnetic field. Gubbiotti et al. explore band structures in meander-shaped magnetic single and bilayers [55] and thereby prepare the way for complex and multi-component 3D magnonic crystals in this Special Topic. Such artificial crystals are expected to support magnon-based data processing [56]. The collection of articles substantiates that nanomagnonics gives rise to unprecedented possibilities for microwave electronics and gets the field ready for the enhancement of its integrated density similar to 3D integrated microelectronic circuits.

**3D Mesoscopic Magnetic Structures and Curvilinear Magnetism**

At present, tailoring of chiral and anisotropic magnetic responses is typically carried out by optimizing materials, either doping a bulk single crystal or adjusting interface properties of thin films and multilayers. A viable alternative to the conventional material screening approach relies on the exploration of the interplay between the geometry of a magnetic object and topology of the magnetic order parameter. Curvilinear magnetism addressees the impact of the geometric curvature on magnetic responses of arbitrarily curved wires and thin films [57]. The lack of inversion symmetry and emergence of the curvature induced anisotropy and DMI are characteristic of curved surfaces, leading to curvature-driven magnetochiral effects and topologically induced magnetization patterning. In this Special Topic, Denis Sheka discusses recent achievements in the field of curvilinear magnetism and outlines perspective research directions [58]. Fundamentals of the magnetisation reversal in Archimedean spirals made of permalloy are addressed by Brajuskovic and Phatak [59]. The effect of the presence of local curvatures in a racetrack on the propagation of noncollinear magnetic textures is discussed by Carvalho-Santos et al. for skyrmions [34] and by Bittencourt et al. for domain walls [60]. Furthermore, the stability, statics and Oersted field induced dynamics of magnetic domain walls and topological defects such as Bloch points in curved nanotubes and nanowires are discussed theoretically by Skoric et al. [25] and experimentally by Schöbitz et al. [26]. With their work, Cheenikundil and Hertel demonstrate a synergy between the field of 3D magnetism and artificial spin ice systems by considering magnetic wireframes shaped as a buckyball [9]. Furthermore, recently, there were exciting proposals on the benefits offered by 3D magnetic architectures for the field of magnonics. This is also reflected in this Special Topic, were Gubbiotti et al. explores the impact of 3D meander shaped objects on the magnonic band structure [55], Salazar-Cardona discusses nonreciprocity of spin waves in nanotubes [53] and Dobrovolskiy et al. addresses spin waves in 3D nanovolcanoes [52]. In addition to their relevance for fundamental research, magnetic 3D structures are of interest for sensor applications in flexible electronics [61] as well as in biology and medicine as discussed by Patil et al. for magnetic nanohelices [62]. In the field of curvilinear magnetism, the vast majority of activities is dedicated to curved ferromagnets. Only very recently, has the focus been put on curvilinear antiferromagnets, where Pylypovskyi et al. discovered homogeneous DMI and weak ferromagnetism in anisotropic achiral antiferromagnetically coupled spin chains [63].

**Micro- and Nanomachines**

The last two decades have seen significant advances in magnetic materials processing, micro- and nanofabrication, and magnetic navigation systems, which has enabled the realization of magnetically driven small-scale machines [64]. These devices, known as magnetic small-scale robots or swimmers, are miniaturized untethered end-effectors capable of moving through fluids using external magnetic fields. To display translational motion, small-scale structures must have a non-reciprocal propulsion mechanism [65]. Depending on the robot's design and material and the applied magnetic field form (e.g., oscillating, rotating, gradient or combinations thereof), several locomotion strategies can be achieved (e.g., rolling, tumbling, wobbling or cork-screwing). Translational motion of magnetic swimmers can be achieved by exploiting forces (magnetic field gradients) or torques. The first strategy results in pulling the magnetic body. The second exploits the fact that magnetic bodies usually exhibit an easy magnetization axis, which aligns parallel to the magnetic field causing magnetic bodies to adopt a specific orientation. Hence, a rotating magnetic field causes a magnetic body to rotate with its easy axis. If this rotation results in a non-reciprocal motion, then the magnetic body will exhibit a net displacement. For example, a helix, when it is rotated around its long axis, will display a cork-screwing translational mechanism. Generally, a magnetic swimmer actuated by rotating magnetic field displays a maximum frequency, known as the step-out frequency. Below this value, the swimmer increases its velocity linearly with increasing frequency as the easy axis rotates synchronously with the rotating magnetic field. Beyond the step-out frequency, the swimmer's velocity will exhibit a nonlinear decline when increasing the frequency because the rotating magnetic field is asynchronous with respect to the swimmer's easy axis.

While the community is currently shifting efforts towards applications (e.g., minimally invasive medicine or water cleaning), there is still a focus on understanding small-scale motion and developing magnetic propulsion mechanisms. In this Special Topic, Patil et al. report on the magnetic manipulation of magnetic nanohelices in an optical bowl - a 2D optical confinement [62]. When the helices are subject to a rotating magnetic field with frequencies higher than 15 Hz, the nanoswimmers are able to escape the 2D trap, while below this frequency, they remain confined in the optical bowl. This approach is interesting for investigating rheological properties of fluids and the performance of swarms. In the same issue, Bachmann et al. explore the opportunities of branching and step-out behaviour in magnetic nonsymmetrical dumbbell-shaped microswimmers with nonlinear velocity–frequency relationships [66]. A branching regime is identified when the swimmers are actuated just above the linear tumbling region. Branching refers to two different stable motion mechanisms with different velocities at the same frequency, in this case, two wobbling behaviours with opposed swimming directions. Upon increasing the frequency, the swimmers display a wobble-like step-out motion. The authors suggest that exploitation of these swimming regimes could be used for better control of swarms, and importantly, show that artificial swimmers can adapt their swimming behaviours to a certain extent in a similar way to that of natural microorganisms.

## ACKNOWLEDGEMENTS

We would like to thank all the authors that have contributed to this Special Topic as well as the journal editors and staff who helped to put this great collection together.